\documentclass[aps,onecolumn,groupedaddress,nofootinbib,amsmath,amssymb]{revtex4}
\usepackage{dsfont}
\usepackage{bm}
\begin{document}
\setcounter{page}{1}
\title[]{Higher-dimensional charged black holes solutions with a nonlinear electrodynamics source}
\author{Mokhtar Hassa\"{\i}ne}\email{hassaine-at-inst-mat.utalca.cl}
\affiliation{Instituto de Matem\'atica y F\'{\i}sica, Universidad de
Talca, Casilla 747, Talca, Chile,} \affiliation{Laboratoire de
Math\'ematiques et de Physique Th\'eorique, Universit\'e de Tours,
Parc de Grandmont, Tours, France.}
\author{Cristi\'an Mart\'{\i}nez}\email{martinez-at-cecs.cl}
\affiliation{Centro de Estudios Cient\'{\i}ficos (CECS),
 Casilla 1469, Valdivia, Chile\\and
Centro de Ingenier\'{\i}a de la Innovaci\'on del CECS (CIN),
Valdivia, Chile.}

%\preprint{{\tiny CECS-PHY-06/25} }

\begin{abstract} We obtain electrically charged black hole solutions of the
Einstein equations in arbitrary dimensions with a nonlinear
electrodynamics source. The matter source is deriving from a
Lagrangian given by an arbitrary power of the Maxwell invariant. The
form of the general solution suggests a natural partition for the
different ranges of this power. For a particular range, we exhibit a
class of solutions whose behavior resemble to the standard
Reissner-Nordstr\"{o}m black holes. There also exists a range for
which the black hole solutions approach asymptotically the Minkowski
spacetime slower than the Schwarzschild spacetime. We have also
found a family of not asymptotically flat black hole solutions with
an asymptotic behavior growing slower than the Schwarzschild (anti)
de Sitter spacetime. In odd dimensions, there exists a critical
value of the exponent for which the metric involves a logarithmic
dependence. This critical value corresponds to the transition
between the standard behavior and the solution decaying to Minkowski
slower than the Schwarzschild spacetime.

\end{abstract}

\maketitle

%%%%%%%%%%%%%%%%%%%%%%
\section{Introduction}
%%%%%%%%%%%%%%%%%%%%%%

One of the most interesting black hole solutions with a matter
source is the Kerr-Newman solution. The Kerr-Newman solution
describes a massive and rotating charged black hole, which is
solution of the Einstein-Maxwell equations in four dimensions. At
the vanishing angular momentum limit, the Kerr-Newman solution
reduces to the Reissner-Nordstr\"{o}m solution. In four dimensions,
the Maxwell theory is conformally invariant and hence, the
derivations of the Reissner-Nordstr\"{o}m or the Kerr-Newman
solutions are considerably simplified since in this case the scalar
curvature vanishes. In higher dimensions, where the Maxwell theory
is not conformally invariant, the derivation of the corresponding
Reissner-Nordstr\"{o}m solution has been done long time ago
\cite{Tangherlini:1963bw}. However, finding the equivalent of the
Kerr-Newman in higher dimensions is still an open problem. Recently,
we have taken seriously the lack of conformal symmetry of the
Maxwell action for dimensions $d>4$ and proposed the following
generalization of the Maxwell action in arbitrary dimensions
\begin{eqnarray}
I_{\cal M}=-\alpha\int
d^dx\,\sqrt{-g}\left(F_{\mu\nu}F^{\mu\nu}\right)^{q}, \label{extM}
\end{eqnarray}
where $F_{\mu\nu}=\partial_{\mu}A_{\nu}-\partial_{\nu}A_{\mu}$ is
the Maxwell tensor, and $\alpha$ is a constant
\cite{Hassaine:2007py}. The interest of considering such action lies
in the fact that in arbitrary dimension $d$, the action $I_{\cal M}$
can enjoy the conformal invariance provided the exponent to be
chosen as $q=d/4$. For this particular choice of the exponent, we
have derived black hole solutions electrically charged with a purely
radial electric field of the conformal nonlinear electrodynamics for
dimensions of multiples of four. The restriction on the dimension is
due to the electromagnetic field Ansatz which imposes the exponent
to be given only by multiples of four in order to deal with real
solutions \cite{Hassaine:2007py}.  A way of escaping from this
restriction is to consider the absolute value of the Maxwell
Lagrangian as it has been done in three dimensions with the
conformal exponent \cite{Cataldo:2000we}. Note that there exists
another conformally invariant extension of the Maxwell action in
higher dimensions for which the Maxwell field is replaced by a
$d/2-$form with $d$ even \cite{Fabris:1991pj}. The black hole
solutions of this theory were discussed in \cite{Bronnikov:1996df}.

In the present paper, we relax the conformal condition and instead,
we consider the Einstein equations in arbitrary dimension with the
nonlinear electrodynamics source (\ref{extM}) without restricting
{\it \`a priori} the exponent $q$. The option of considering
nonlinear electrodynamics models as sources of the Einstein equation
is motivated by the fact that nonlinear electrodynamics models
provide excellent laboratories for the construction of black hole
solutions with interesting properties, for instance, regular black
holes \cite{regularBH}. Also, in the last years, a number of papers
have been dedicated to find and analyze different solutions in
gravitational theories with nonlinear electrodynamics sources,
including, in some cases, additional matters fields \cite{NLED}.

The plan of the paper is organized as follows. In Section
\ref{ecuaciones}, after presenting the field equations, we determine
the sign of the coupling constant $\alpha$ in terms of the exponent
$q$ in order to satisfy the energy conditions. We derive the most
general static and spherically symmetric solution with a purely
radial electric field in Sec. \ref{soluciones}. The general solution
contains two integration constants and it is possible to find event
horizons, which also depend on $q$ and the spacetime dimension.
Moreover, as it is expected, the two integration constants of the
general solution can be related with the mass and electric charge
and  the higher-dimensional generalization of the
Reissner-Nordstr\"{o}m solution is obtained as the exponent $q$ is
set to $1$. As it is shown along of this section, the general
solution suggests a natural partition for the different ranges of
the exponent $q$.

First of all, there is an excluded region corresponding to the
values $q\in(0,1/2)$ for which the scalar curvature of the solutions
diverges at infinity. For $q\in\,(1/2,(d-1)/2)\cap
\tilde{\mathds{Q}}\,$, where the set $\tilde{\mathds{Q}}$ denotes
the rational number with odd denominator (\ref{set}), the solution
has a behavior similar to the standard Reissner-Nordstr\"{o}m
solution. By similar we mean that asymptotically the charge
contribution in the metric goes to zero faster than the mass
contribution. At the opposite and surprisingly enough, black hole
solutions which go asymptotically to the Minkowski spacetime slower
than the Schwarzschild spacetime are also exhibited for $q>(d-1)/2$
and for $q<-1/(d-4)$ with $q\in \tilde{\mathds{Q}}$. In four
dimensions, these solutions only exist for $q>3/2$ with $q\in
\tilde{\mathds{Q}}$. For the remaining values of the exponent, i. e.
$q\in (-1/(d-4),0)\cap \tilde{\mathds{Q}}$, the black hole solutions
are not asymptotically flat and, their asymptotic behavior is shown
to grow slower that the Schwarzschild (anti) de Sitter spacetime. In
four dimensions, these not asymptotically flat black hole solutions
are exhibited for all non positive exponents belonging to the set
$\tilde{\mathds{Q}}$.

The different black hole solutions are classified, depending on
their asymptotic behaviors, among these four different classes. The
transitions between these different classes of solutions are
operated at the critical values $q_{c_1}=(d-1)/2$ and
$q_{c_2}=-1/(d-4)$, and it is clear that only in odd dimensions
these critical values belong to the set $\tilde{\mathds{Q}}$. Thus,
an analysis of the solutions at the critical exponents is done in
odd dimensions. At the first critical exponents $q_{c_1}$ which
corresponds to the transition between the standard behavior and the
solution decaying to Minkowski spacetime slower than the
Schwarzschild spacetime, it is shown that the metric involves a
logarithmic dependence and the expression of the electric potential
is proportional to $\ln r$ independently of the odd dimensions. At
the other critical value $q_{c_2}$, the transition is smooth and
corresponds to the region where the metric function asymptotically
goes to a constant (not necessarily equal to 1) proportional to the
black hole charge. After this detailed study of all the cases, the
Sec. \ref{discusion} is concerned with the comments and the possible
extensions of the present work.

%%%%%%%%%%%%%%%%%%%%%%%%%
\section{Field equations} \label{ecuaciones}
%%%%%%%%%%%%%%%%%%%%%%%%%

We consider the following action in $d>2$ dimensions
\begin{eqnarray} I[g_{\mu\nu},A_{\mu}]=\int
d^{d}x\,\sqrt{-g}\Big[\frac{R}{2\kappa}-\alpha\left(F_{\mu\nu}F^{\mu\nu}\right)^{q}\Big],
\label{action}
\end{eqnarray}
where $q$ is a rational number that will be fixed later, $R$ is the
scalar curvature, and $\kappa >0$ is the gravitational constant. The
field equations obtained by varying the metric and the gauge field
$A_{\mu}$ read respectively
\begin{subequations}
\label{eqs}
\begin{eqnarray}
&&G_{\mu \nu }=4\kappa\alpha\left[q
F_{\mu\rho}F_{\nu}^{\,\,\,\rho}\,F^{q-1}-
\frac{1}{4}g_{\mu\nu}\,F^{q}\right], \label{EE} \\
&&\frac{1}{\sqrt{-g}}\partial_{\mu}\left(\sqrt{-g}\,F^{\mu\nu}\,F^{q-1}\right)=0,\label{EE1}
\end{eqnarray}
\end{subequations}
where $F$ is the Maxwell invariant
$F=F_{\alpha\beta}F^{\alpha\beta}$.

We are looking for a static and spherically symmetric spacetime
geometry whose line element is given by
\begin{eqnarray}
ds^2=-N^2(r)f^2(r)\,dt^2+\frac{dr^2}{f^2(r)}+r^2\,d\Omega_{d-2}^2,
\label{spacetimegeo}
\end{eqnarray}
where $d\Omega_{d-2}^2$ is the line element of the
$(d-2)-$dimensional sphere. In addition, we only consider a purely
radial electric Ansatz for the electromagnetic field which means
that the only non-vanishing component of the Maxwell tensor is given
by $F_{tr}$. As a direct consequence, the Maxwell invariant
$F=-2(F_{tr})^2$ is negative and hence, the exponent $q$ can only be
an integer or a rational number with odd denominator. Hence, in
order to deal with real solutions, the exponent $q$ is restricted to
be an element of the following set \footnote{ Note that $\mathds{Z}
\subset \tilde{\mathds{Q}}.$}
\begin{equation}
\tilde{\mathds{Q}}=\left\{\frac{n}{2p+1},\quad
(n,p)\in\mathds{Z}\times\mathds{Z} \right\}. \label{set}
\end{equation}
Taking the trace of the Einstein equations, the scalar curvature is
expressed in terms of the Maxwell invariant $F$ as
\begin{eqnarray}
R=2\kappa\alpha\frac{(4q-d)}{(2-d)}F^q. \label{curvature}
\end{eqnarray}
Using this expression for the scalar curvature, the Einstein
equations can be written as
\begin{subequations}
\begin{eqnarray}\label{Einsteintensor}
&&G^{t}_{t}=-\frac{((N f)' f)'}{N}-(d-2)\frac{(N f)' f}{r N}-\kappa\alpha\frac{(4q-d)}{(2-d)}F^q= \kappa\alpha (2q-1)F^q\label{Et1} \\
&&G^{r}_{r}=-\frac{((N f)' f)'}{N}-(d-2)\frac{f' f}{r }-\kappa\alpha\frac{(4q-d)}{(2-d)}F^q= \kappa\alpha (2q-1)F^q\label{Et2} \\
&&G^{\theta_i}_{\theta_i}=-\frac{(N f)' f}{r N}-\frac{f' f}{r
}+\frac{d-3}{r^2}(1-f^2)-\kappa\alpha\frac{(4q-d)}{(2-d)}F^q=-\kappa\alpha
F^q\label{Et3}
\end{eqnarray}
\end{subequations}
where $\theta_i$, with $i=1,\cdots, (d-2)$, are the angular
coordinates and the prime denotes derivative with respect to the
radial coordinate $r$. Subtracting Eqs. (\ref{Et1}) and (\ref{Et2})
we obtain that
\begin{equation}
-(d-2)\frac{N' f^2}{r N}=0,
\end{equation}
hence  $N(r)$ is a constant, which can be set to 1 without
loss of generality.

Before studying in details the field equations, we first specify the
sign of the coupling constant $\alpha$ in term of the exponent $q$
in order to ensure a physical interpretation of our future
solutions. In fact, the sign of the coupling constant $\alpha$ in
the action (\ref{action}) can be chosen such that the energy
density, i.e. the $T^{\hat{0}\hat{0}}$ component of the
energy-momentum tensor in the ortonormal frame, is positive
$$ T^{\hat{0}\hat{0}}= -\kappa\alpha (2q-1)F^q >0.$$
This condition selects two branches depending on the range of the
exponent $q$,
\begin{equation}
\left\lbrace
\begin{array}{l}
\mbox{sgn}(\alpha)=-(-1)^q\qquad\mbox{for}\quad q>1/2,\\
\mbox{sgn}(\alpha)=(-1)^q\qquad\quad\mbox{for}\quad q<1/2,
\end{array}
\right. \label{sgn}
\end{equation}
while the case $q=1/2$ is excluded by the condition (\ref{set}).

%%%%%%%%%%%%%%%%%%%%%%%%%%%%%%%%%%%%%%%%%%%%%%%%%%%%
\section{Electrically charged black hole solutions}  \label{soluciones}
%%%%%%%%%%%%%%%%%%%%%%%%%%%%%%%%%%%%%%%%%%%%%%%%%%%%
 Let us now derive the most general solution of the static and
spherically symmetric Einstein equations (\ref{eqs}) with a purely
radial electric field. As said previously, the substraction of the
equations (\ref{Et1}) and (\ref{Et2}) implies that the metric
function $N$ can be set to $1$. As a direct consequence, the
generalized Maxwell equation (\ref{EE1}) reads
\begin{equation}
\partial_{r}\left( r^{d-2} \,F^{r t}\,(F_{r t})^{2q-2}\right)=0,
\end{equation}
which
 can easily be solved
yielding to $F_{tr}\propto{r^{\frac{2-d}{2q-1}}}$. Hence, the
Einstein equations become differential equations of the only metric
function $f$ and their integration yields to the following solution
\begin{subequations}
\label{sol}
\begin{eqnarray}
&&F_{tr}=\frac{C}{r^{\frac{d-2}{2q-1}}}, \label{1} \\
&&f^2(r)=1-\frac{A}{r^{d-3}}-\frac{2\kappa\alpha(-1)^q
C^{2q}\,2^q\,(2q-1)^2}{(d-2)(d-2q-1)r^{\frac{2(qd-4q+1)}{2q-1}}},\label{2}\\
&& N(r)=1, \label{3}
\end{eqnarray}
\end{subequations}
where $A$ and $C$ are two integration constants proportional to the
mass and the electric charge. Various comments can be made
concerning the structure of this solution. Firstly, as it can be
seen from the metric function (\ref{2}), the odd dimensions $d=2n+1$
with $q=n$ requires a special analysis. In fact, as it will be shown
below this critical value corresponds to the transition between the
standard behavior and the solution decaying to Minkowski slower than
the Schwarzschild spacetime. The solutions derived here reduce to
the higher-dimensional generalizations of the Reissner-Nordstr\"{o}m
solution \cite{Tangherlini:1963bw} for $q=1$ as it should be since
this limiting case is not singular. In the conformal situation, i.
e. $q=d/4$, the solutions (\ref{sol}) exist only for dimensions of
multiples of four because of the restriction (\ref{set}), and our
previous results \cite{Hassaine:2007py} are recovered. It is also
interesting to note that the Schwarzschild (anti) de Sitter solution
can be obtained from the expression (\ref{2}) by putting $q=0$. This
result is not surprising since at the level of the action
(\ref{action}), putting $q=0$ is equivalent of considering the
Einstein action with a cosmological constant $\Lambda$ given in term
of the coupling constant $\alpha$ as $\Lambda= \kappa \alpha$.

We are interested in finding solutions with event horizons. These
horizons should be hide curvature singularities, and hence solutions
having singularities at infinity will be ruled out and only
curvature singularities surrounded by an event horizon are allowed.
For the solution (\ref{sol}), the scalar curvature has a single
singularity at the origin $r=0$ if $q\in\tilde{\mathds{Q}}$ and $q
>1/2$ or $q\le 0$\textbf{,} while the range $0<q<1/2$ is excluded since the scalar
curvature diverges at the infinity. The analysis of the Kretchsmann
invariant, $K=R_{\mu \nu \lambda \rho} R^{\mu \nu \lambda \rho}$,
leads to the same conditions on $q$ in order to satisfy the
regularity condition at infinity. Moreover, it is easy to show that
the curvature regularity condition as $r\to \infty$, in addition to
the positive energy condition, ensure that the weak energy condition
is satisfied.

The existence of the horizons for the solutions (\ref{2}) must be
done carefully because of the presence of various parameters as the
constants appearing in the action, the dimension $d$ and the
exponent $q$, as well as the integration constants $A$ and $C$.
Hence, for clarity we first rewrite the metric function (\ref{2})
schematically as
\begin{eqnarray}
f^2(r)=1-\frac{A}{r^{d-3}}+\frac{B}{r^{\beta}} \label{beta}
\end{eqnarray}
where we have defined
\begin{eqnarray}
B=-\frac{2\kappa\alpha(-1)^q
C^{2q}\,2^q\,(2q-1)^2}{(d-2)(d-2q-1)},\qquad
\beta=\frac{2(qd-4q+1)}{2q-1}. \label{parameters}
\end{eqnarray}
The form of the metric solution suggests two natural ranges
concerning the exponent $\beta$, namely whether if $1/r^{\beta}$
goes faster or not to zero than the Schwarzschild potential
$1/r^{d-3}$, i.e. $\beta>d-3$ or $0<\beta<d-3$, respectively. The
remaining possibilities are $\beta=d-3$ which corresponds to the
critical exponent $q=(d-1)/2$ and $\beta\leq 0$ for which the metric
is not asymptotically flat. All these different possibilities are
now analyzed in detail.

%%%%%%%%%%%%%%%%%%%%%%%%%%%%%%%
\subsection{Case $\beta>d-3$}
%%%%%%%%%%%%%%%%%%%%%%%%%%%%%%%
The case $\beta>d-3$ is similar to the standard
Reissner-Nordstr\"{o}m case in the sense that for large $r$, the
metric behaves like the Schwarzschild metric. The condition
$\beta>d-3$ imposes the exponent $q$ to be in the following range
$q\in \tilde{\mathds{Q}}\,\,\cap\,\,(1/2, (d-1)/2)$ which in turn
implies that the electric field (\ref{1}) vanishes at infinity. On
the other hand, the positivity of the energy density (\ref{sgn})
imposes the constant $B$ to be positive. In this case, in order to
have real roots for the metric function $f^2(r)$, the constant $A$
must be positive and the constant $B$ must be chosen in the
following range
\begin{eqnarray}
0<B<(d-3)\left(\frac{A}{\beta}\right)^{\frac{\beta}{d-3}}(\beta+3-d)^{\frac{\beta+3-d}{d-3}}.
\end{eqnarray}
Under these conditions, we have two roots localized at
$r_{-}\in(0,b)$ and $r_{+} \in (b,\infty)$ where
$$
b=\left(\frac{A}{\beta}(\beta+3-d)\right)^{\frac{1}{d-3}}.
$$
An extreme black hole can also be obtained if $A$ is positive and
the constant $B$ given by
$$
B=(d-3)\left(\frac{A}{\beta}\right)^{\frac{\beta}{d-3}}(\beta+3-d)^{\frac{\beta+3-d}{d-3}}.
$$

%%%%%%%%%%%%%%%%%%%%%%%%%%%%%%%
\subsection{Case $0<\beta<d-3$}
%%%%%%%%%%%%%%%%%%%%%%%%%%%%%%%
The case $0<\beta<d-3$ is compatible with exponents $q$ given by
$$
q\in\tilde{\mathds{Q}}\,\,\cap\,\,(\frac{d-1}{2},\infty)\qquad
\mbox{or}\qquad q\in\tilde{\mathds{Q}}\,\,\cap\,\,(-\infty,
\frac{-1}{d-4})
$$
This situation is surprisingly enough since the Schwarzschild
potential $1/r^{d-3}$ goes faster to zero than the potential
proportional to the electric charge. The energy condition implies
that the constant $B$ is always negative, and hence if the constant
$A$ is positive, there is a single root
$$
r\in]r_0,\infty[,\qquad\qquad
r_0=\left(\frac{-B(d-3-\beta)}{d-3}\right)^{1/\beta}
$$
On the other hand if the constant $A<0$, there are two roots if
$A>A_0$  where
\begin{eqnarray}
A_0=-\frac{\beta}{d-3-\beta}\left(\frac{-B(d-3-\beta)}{d-3}\right)^{(d-3)/\beta}
\label{Ao}
\end{eqnarray}
and a double root if $A=A_0$.

The emergence of black hole solutions for $0<\beta<d-3$ is
interesting by itself since the solutions approach asymptotically
the Minkowski spacetime slower than the Schwarzschild spacetime.
Such asymptotic behavior has been observed in the case of massive
scalar fields minimally coupled to anti de Sitter gravity with a
potential \cite{Henneaux:2006hk}.

Another amusing fact is concerned with the odd dimensions given by
$d=7+4p$ where $p\in\mathds{N}$ and $q=-(2+p)/(1+2p)$ for which the
metric solution becomes
$$
f^2(r)=1+\frac{\kappa\,\alpha\,
(-1)^{q+1}\,C^{2q}\,2^{q-1}}{(p+1)\,(2p+1)\,r^{2p+2}}-\frac{A}{r^{2(2p+2)}}.
$$
It is clear from this expression that, up to the physical
interpretation of the integration constants, this metric function
corresponds to Reissner-Nordstr\"{o}m  metric function in odd
dimension $\tilde{d}=2p+5=(d+3)/2$. This analogy is clearly not
respected by the electric field since the Maxwell electric field in
the nonlinear case is given by $F_{tr}=C\,r^{2p+1}$ and does not
correspond to the Reissner-Nordstr\"{o}m solution in dimension
$\tilde{d}=2p+5$.

%%%%%%%%%%%%%%%%%%%%%%%%%%%%%%%
\subsection{Case $\beta=d-3$}
%%%%%%%%%%%%%%%%%%%%%%%%%%%%%%%
The limiting case $\beta=d-3$ corresponds to the change of phase
between the solution which resembles to the standard
Reissner-Nordstr\"{o}m solution and the one for which the asymptotic
decaying to Minkowski spacetime is slower than the Schwarzschild
spacetime. This case is compatible with the values of the exponent
$q$ given by $q=(d-1)/2$, and because of the restriction
(\ref{set}), the exponent can only be a positive integer $q\geq 1$.
This implies that for the critical exponent $q=(d-1)/2$, we are only
concerned with odd dimensions $d=2q+1$. This value is critical in
the sense that the constant $B$ of the solution (\ref{parameters})
is singular at this value. In this case, the integration of the
Einstein equations for a static and spherically symmetric spacetime
in presence of a purely radial electric field yields
\begin{subequations}
\label{sol2}
\begin{eqnarray}
&&F_{tr}=\frac{C}{r}, \label{11} \\
&&f^2(r)=1-\frac{A}{r^{2q-2}}+\kappa\alpha(-1)^q
2^{q+1}C^{2q}\frac{\ln r}{r^{2q-2}},\label{21}
\end{eqnarray}
\end{subequations}
where $A$ and $C$ are two constants of integration. For a constant
$A>1$, the existence of horizons for the logarithmic solution
(\ref{sol2}) is always ensured. In the remaining case $A\leq 1$, the
metric function (\ref{21}) can have roots provided one fixes
properly the constant $C$. Some comments can be added concerning
this particular solution. The three-dimensional case corresponds to
the standard Maxwell case, i.e. $q=1$, and the expression
(\ref{sol2}) reduces to the three-dimensional Einstein-Maxwell
solution \cite{Banados:1992wn}. The emerging of a logarithmic
function in the metric is similar to the case studied in
\cite{Henneaux:2004zi}, where the slow asymptotic fall-off produced
by these logarithmic branches appear in the context of a
self-interacting scalar field whose mass saturates the
Breitenlohner-Freedman bound, minimally coupled to Einstein gravity
with a negative cosmological constant in $d >2$ dimensions. It is
also interesting to remark that for any odd dimensions $d>3$, the
solution approaches to Minkowski space in the asymptotic region, but
it is not asymptotically flat due to the presence of a logarithmic
term. Also, it is remarkable that independently of the value of the
exponent $q>1$, the expression of the electric field does not depend
on the dimension. This means that in any odd dimensions $d=2n+1$ and
with $q=n$, the electric field behaves like the Maxwell field in
three dimensions. A same analogy has been observed in dimensions
$d=4n$ with $q=n$, where the expression of the electric field is the
same than the four-dimensional Coulomb field \cite{Hassaine:2007py}.
The hierarchical character of the electric field of the solutions,
as the exponent $q$ takes integer values, can be generalized in the
following sense. In arbitrary dimension $\tilde{d}$, the electric
field of the Reissner-Nordstr\"{o}m solution goes like
$F_{tr}\propto 1/r^{\tilde{d}-2}$ \cite{Tangherlini:1963bw}. From
the generic solution (\ref{1}), it is clear that such behavior can
be obtained in dimensions $d\geq\tilde{d}$ such that
$$
d=(\tilde{d}-2)(2q-1)+2,
$$
with the exponent $q\in\mathds{N}$.

%%%%%%%%%%%%%%%%%%%%%%%%%%%%%%%
\subsection{Case $\beta <0$}
%%%%%%%%%%%%%%%%%%%%%%%%%%%%%%%
It remains to study the case for which the metric solution
(\ref{sol}) is not asymptotically flat, i.e. $\beta <0$. First of
all, the condition $\beta<0$ is consistent only with $q\in
(-1/(d-4),1/2)$ and, since we are not considering the solutions
having singularities at the infinity, this interval is restricted to
be $q\in (-1/(d-4),0)$. On the other hand, the positivity of the
energy density (\ref{sgn}) implies that the constant $B$ in
(\ref{parameters}) is negative. In this situation, for a constant
$A>0$, the metric function will have two roots provided that the
constant $A$ is bounded as follows
$$
A<A_0,\qquad
A_0=\frac{-\beta}{d-3-\beta}\left(\frac{-(d-3)}{B(d-3-\beta)}\right)^{(3-d)/\beta}
$$
while a double root is obtained for $A = A_0$, however, this does
not lead to an extreme black hole. In the case for which $A<0$,
there is a single root localized in the following interval
$]r_0,\infty[$ where
$$
r_0=\left(\frac{-(d-3)}{B(d-3-\beta)}\right)^{-1/\beta}.
$$

Another interesting fact is that the metric function behaves
asymptotically like $f(r)^2\sim r^{-\beta}$ with $0<-\beta<2$ which
in turn implies that the metric function goes to infinity slower
than the Schwarzschild (anti) de Sitter spacetime. The only option
for the metric function to go faster to infinity than the
Schwarzschild (anti) de Sitter geometry is for $q\in (0,1/2)$ which
is precisely the forbidden region where the solution has a naked
singularity.

%%%%%%%%%%%%%%%%%%%%%
\section{Discussion} \label{discusion}
%%%%%%%%%%%%%%%%%%%%%
In this paper, we have seen the influence of considering arbitrary
power of the Maxwell Lagrangian as a source of the Einstein
equations for a static and spherical symmetric spacetime geometry.
We have obtained the most general black hole solutions of the
Einstein equations for static and spherically symmetric spacetime
with a purely radial electric field. The general solutions depend on
the dimension, the exponent and reduce to the Reissner-Nordstr\"{o}m
solutions as the exponent $q=1$. We have exhibited some interesting
properties of these black hole solutions. The range of the exponent
$q$ has been divided in four parts depending on the asymptotic
behavior of the solution. For example, we have derived black hole
solutions that go asymptotically to the Minkowski spacetime faster
or slower than the Schwarzschild solution. We have also found a
range for which the black hole solutions are not asymptotically flat
generalizing the standard Schwarzschild (anti) de Sitter solution
without requiring the introduction of a cosmological constant. There
also exists a critical value of the exponent only in odd dimensions
for which the metric solution involves a logarithmic dependence and
the behavior of the electric field does not depend on the dimension
and is given by the standard three-dimensional Maxwell field.

It will be interesting to realize a similar analysis in the case of
a static and axisymmetric Ansatz for the metric and to look for
rotating charged black hole solution. There are essentially two
interesting options to explore in this perspective. The first one is
to consider the four-dimensional problem with an arbitrary power of
the Maxwell Lagrangian in order to obtain a metric generalizing the
Kerr-Newman geometry, and also to see the influence of the exponent
on the asymptotic behavior of the metric. Mathematically, this work
is highly non trivial because the field equations are complicated
and, in contrast with the Einstein-Maxwell case, the matter source
in the nonlinear case is not longer conformally invariant. The other
option is to consider the same problem in higher dimensions but with
a conformal nonlinear source, i. e. $q=d/4$. In the static and
spherical case and for a purely electric field, we have already
solved this problem \cite{Hassaine:2007py}. In the same spirit, we
can also look for other kind of solutions as for example the charged
Taub-NUT solutions in even dimensions \cite{Brill}.

In the metric solution (\ref{sol}), there appear two constants of
integration that are proportional to the mass and the electric
charge. It will be interesting to identify precisely the mass and
the electric of the black hole solutions. In the case of solutions
which are asymptotically flat, the use of the Hamiltonian action may
provide a simple manner of identifying the black hole mass and
charge which is not necessarily the case for the solutions with
$\beta<0$.

Another interesting aspect to explore will be the stability of the
solutions derived here. In this context, the authors of Ref.
\cite{Moreno:2002gg} have derived conditions on a class of the
electromagnetic Lagrangians to ensure the linear stability of black
hole configurations.

Finally, it is also be desirable to study the geometric properties,
the causal structures as well as the thermodynamics properties of
the black hole solutions derived here. In particular, these
questions may be of interest in the case of the solutions derived
for which the spacetimes are not asymptotically flat.

\bigskip

\acknowledgments  We thank Eloy Ay\'on-Beato and Jorge Zanelli for
useful discussions. This work is partially funded by FONDECYT
(Chile) grants 1051084, 1060831, 1051064, 1051056, 1061291, 1071125,
1085322. The Centro de Estudios Cient\'{\i}ficos (CECS) is funded by
the Chilean Government through the Millennium Science Initiative and
the Centers of Excellence Base Financing Program of Conicyt. CECS is
also supported by a group of private companies which at present
includes Antofagasta Minerals, Arauco, Empresas CMPC, Indura,
Naviera Ultragas and Telef\'onica del Sur. CIN is funded by Conicyt
and the Gobierno Regional de Los R\'{\i}os.

%%%%%%%%%%%%%%%%%%%%%%%%%%%

\end{document}